\documentclass[twocolumn,prl,10pt,superscriptaddress]{revtex4-1}
\usepackage[dvipsnames,usenames]{color}
\usepackage{graphicx}
\usepackage{dcolumn}
\usepackage{bm}
\usepackage{slashed}
\usepackage{amssymb}
\usepackage{amsmath}

\usepackage{xr}
\usepackage{color}
\usepackage[colorlinks]{hyperref}

\begin{document}

	\title{Optimal control of stimulated Raman adiabatic passage in a superconducting qudit}
	
	\author{Wen Zheng}
	\thanks{These authors contributed equally to this work.}
	\affiliation{National Laboratory of Solid State Microstructures, School of Physics,
		Nanjing University, Nanjing 210093, China}
	\author{Yu Zhang}
	\thanks{These authors contributed equally to this work.}
	\affiliation{National Laboratory of Solid State Microstructures, School of Physics,
		Nanjing University, Nanjing 210093, China}

	\author{Yuqian Dong}
	\affiliation{National Laboratory of Solid State Microstructures, School of Physics,
		Nanjing University, Nanjing 210093, China}
		
	\author{Jianwen Xu}
	\affiliation{National Laboratory of Solid State Microstructures, School of Physics,
		Nanjing University, Nanjing 210093, China}

	\author{Zhimin Wang}
	\affiliation{National Laboratory of Solid State Microstructures, School of Physics,
		Nanjing University, Nanjing 210093, China}

	\author{Xiaohan Wang}
	\affiliation{National Laboratory of Solid State Microstructures, School of Physics,
		Nanjing University, Nanjing 210093, China}
	\author{Yong Li}
	\affiliation{National Laboratory of Solid State Microstructures, School of Physics,
		Nanjing University, Nanjing 210093, China}

	\author{Dong Lan}
	\affiliation{National Laboratory of Solid State Microstructures, School of Physics,
		Nanjing University, Nanjing 210093, China}
	\author{Jie Zhao}
	\affiliation{National Laboratory of Solid State Microstructures, School of Physics,
		Nanjing University, Nanjing 210093, China}
	\author{Shaoxiong Li}
	\affiliation{National Laboratory of Solid State Microstructures, School of Physics,
		Nanjing University, Nanjing 210093, China}
	
	\author{Xinsheng Tan}
	\email{tanxs@nju.edu.cn}
	\affiliation{National Laboratory of Solid State Microstructures, School of Physics,
		Nanjing University, Nanjing 210093, China}
	
	\author{Yang Yu}
	\email{yuyang@nju.edu.cn}
	\affiliation{National Laboratory of Solid State Microstructures, School of Physics,
		Nanjing University, Nanjing 210093, China}

	\date{\today}

	\begin{abstract}    
		{
			
			Stimulated Raman adiabatic passage (STIRAP) is a widely used protocol to realize high-fidelity and robust quantum control in various quantum systems.  
			However, further application of this protocol in superconducting qubits is limited by population leakage caused by the only weak anharmonicity.
			Here we introduce an optimally controlled shortcut-to-adiabatic (STA) technique to speed up the STIRAP protocol in a superconducting qudit. 
			By modifying the shapes of the STIRAP pulses, we experimentally realize a fast ($32$ ns) and high-fidelity ($0.996 \pm 0.005$) quantum state transfer.
			In addition, we demonstrate that our protocol is robust against control parameter perturbations. 
			Our stimulated Raman shortcut-to-adiabatic passage transition provides an efficient and practical approach for quantum information processing.
			}    
	\end{abstract}
	

	\maketitle

	\textbf{Introduction.} 
	
	Adiabatic passage techniques have been widely used to achieve reliable quantum control
	in quantum information processing. 
	Among these techniques, stimulated Raman adiabatic passage (STIRAP) has achieved great success in physics, chemistry and beyond, 
	since it was introduced by Gaubatz et. al. in 1990\cite{GaubatzPopulationswitching,
		1990JChPh..92.5363G,
		vitanov2001laser,
		RevModPhys.79.53,
		shore2008coherent,
		shore2013pre, 
		bergmann2015perspective,
		RevModPhys.89.015006}.
	In recent years, this approach is applied to quantum computation and quantum 
	communication in superconducting circuits\cite{
		kumar2016stimulated, 
		xu2016coherent, 
		vepsalainen2019superadiabatic, 
		yang2019realization,
		niu2019quantum,
		liu2020coherent,
		li2021coherent}. 
		In addition, high-fidelity state transfer has been demonstrated in communication systems with multiple quantum nodes\cite{
		PhysRevLett.124.240502}.
	In a STIRAP scheme, the system Hamiltonian evolves adiabatically, 
	so that its evolution trajectory is insensitive to the loss channel and noise\cite{
		PhysRevA.89.013831,
		PhysRevA.102.023715, 
		PhysRevA.88.013807}.
	This feature makes the approach important for future implementations of deterministic multi-qubit entanglement in 
	quantum communication or computing networks based on superconducting circuits.
	
	In general, the implementation of STIRAP is constrained by the strict adiabatic condition\cite{
			kato1950adiabatic,
			PhysRevA.40.6741,
			1990JChPh..92.5363G,
			kuhn1992population,
			Berry_2009,
			demirplak2008consistency,
			PhysRevLett.111.100502}, 
	which is usually not feasible for superconducting qubits due to their fast decoherence caused by strong coupling with the environment\cite{PhysRevA.76.042319, RevModPhys.85.623, gu2017microwave, RevModPhys.93.025005}. 
	To make the STIRAP approach more practical, it is desirable to speed up the adiabatic procedure to counter the decoherence effect. 
	For adiabatic quantum processes including STIRAP,
	in addition to combining composite pulses to enhance robustness while ensuring high fidelity\cite{LEVITT198661, PhysRevLett.113.043001, PhysRevA.87.043418, PhysRevA.98.053413, PhysRevA.100.023415, PhysRevA.103.052612},
	various shortcut-to-adiabatic (STA) protocols have been theoretically and experimentally studied\cite{
		PhysRevLett.85.1626, torrontegui2013shortcuts, RevModPhys.91.045001}, including counter-diabatic (CD) driving\cite{
			UNANYAN199748, PhysRevA.59.3751, petiziol2020optimized,
			Berry_2009,
			demirplak2008consistency,
			PhysRevLett.111.100502},
	invariants and scaling laws\cite{
			xu2019invariant,
			2020QuIP...19...83F,
			han2021realization,
			song_robust_2021},
	variational methods\cite{KOLODRUBETZ20171, sels2017minimizing},
	and fast forward\cite{lewis1969exact, PhysRevLett.104.063002}. 
	However, introducing these STA protocols in STIRAP (known as STIRSAP) may not be a good solution in transmon qubits due to weak anharmonicity. 
	For instance, with CD driving, to suppress the non-adiabatic excitation from eigenstates, a coupling between the initial state and the target state is introduced\cite{vepsalainen2019superadiabatic,yang2019realization}. 
	However, this modification to the Hamiltonian will lead to unwanted population leakage in the adiabatic procedure\cite{PhysRevA.76.042319}, prohibiting further improvement of transfer fidelity.
	It is possible to optimize the parameters of the driving microwave to suppress unwanted transitions for two-level systems. 
	One can also use the derivative removal by adiabatic gate (DRAG) approach \cite{PhysRevLett.103.110501, PhysRevA.83.012308} in a single qubit gate, but this procedure is not suitable for general three-level systems. 
	
	In this article, we improve the original STIRAP approach by experimentally demonstrating high-fidelity quantum transfer using STIRAP with optimized CD driving.  
	By mapping the three-level system to a two-level system, we add an additional CD driving term to the original Hamiltonian.  We can then neglect the other dynamical coupling terms\cite{
		PhysRevA.94.063411,
		du2016experimental,
		PhysRevLett.116.230503}.
	Moreover, considering the effect of driving pulses on the transmon energy levels, 
	we adopt the covariance matrix adaptation evolution strategy (CMA-ES)\cite{hansen2016cma} 
	algorithm, which is a derivative-free evolution strategy, to optimize the leakage of the population on a non-computational basis. Our protocol provides a robust, high-fidelity method for fast quantum state transition. 
	
	~\\ 
	\textbf{Results.}
	
\begin{figure}
		\begin{minipage}[b]{0.5\textwidth}
			\centering
			\includegraphics[width=8.5cm]{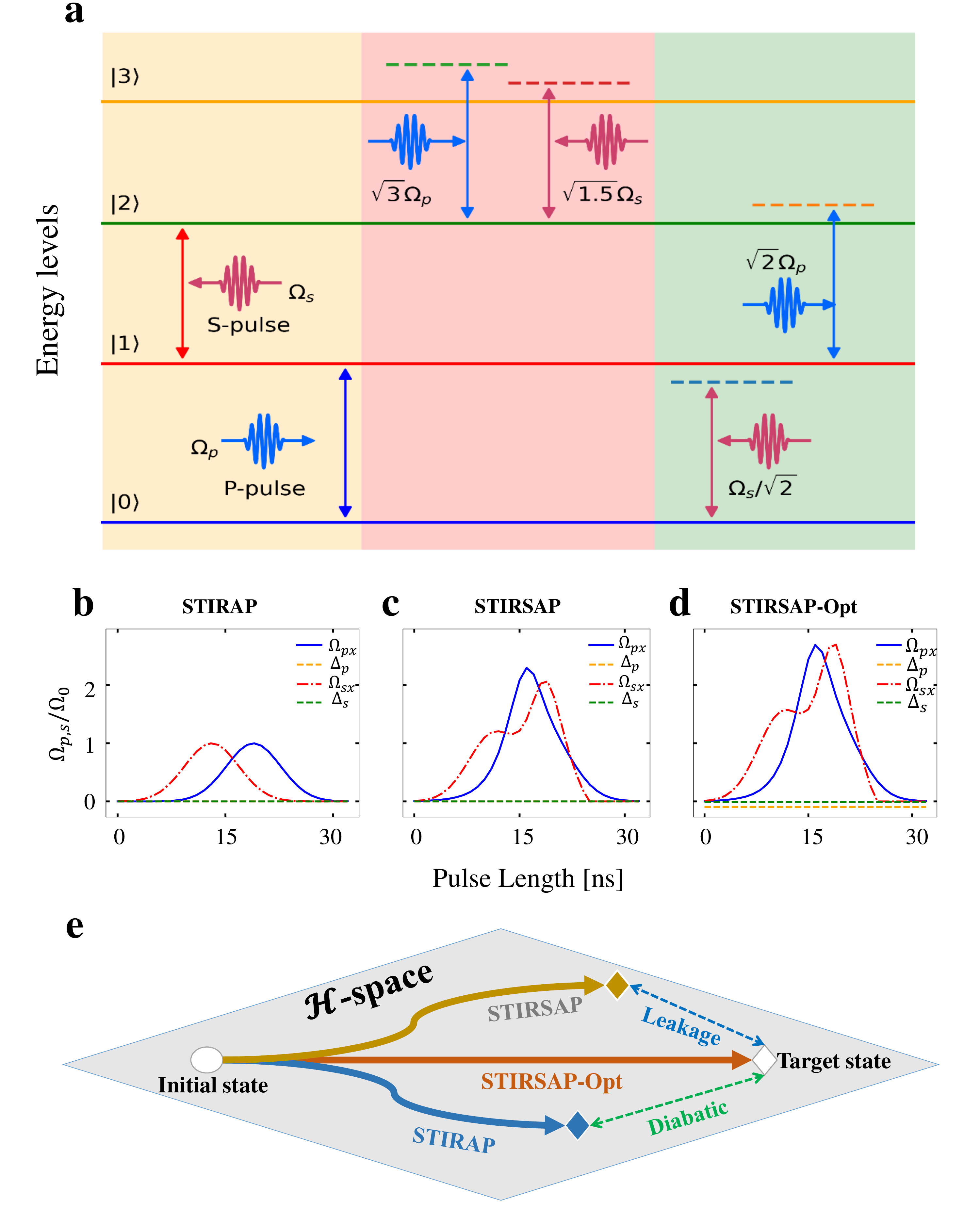}
		\end{minipage}
		\caption{
			\textbf{Experimental schematic diagram.}
			\textbf{a}$\,$ Level diagram of the transmon 
			driven by two external pulses with time-dependent amplitudes.
			In the first (orange) zone, 
			the red (blue) pulse is the resonant pulse between 
			$|0\rangle$ ($|1\rangle$)
			and
			$|1\rangle$ ($|2\rangle$),
			which is called the P-pulse (S-pulse).
			In the second (red) and third (green) zones, the red (blue) pulse is spurious coupling
			to other levels due to weak anharmonicity in the transmon.
			\textbf{b-e}$\,$ Overview of different passages. As shown in \textbf{e}, 
			in a multi-level system with weak anharmonicity,
			the conventional STIRAP (blue line) and STIRSAP (orange line) cannot achieve 
			high-fidelity state transfer control due to the diabatic process and leakage
			when the evolution time is shortened, 
			while STIRSAP-Opt (red line) can realize high fidelity. 
			The envelopes of the three pulses from \textbf{b} to \textbf{d} are STIRAP, STIRSAP and STIRSAP-Opt.
			\label{fig:FIG1}
		}
	\end{figure}

	\textit{Shortcut-to-adiabatic approach.} Firstly, we consider the standard STIRAP approach, in which the Hamiltonian is written as  
	$H=\frac{1}{2}[\Omega_p(t)|0\rangle \langle 1| 
	+ \Omega_s(t)e^{-i\phi}|1\rangle \langle 2| 
	+ h.c.]$ ($\hbar$ is set as 1).
	Its eigenstates include a dark state 
	$|D\rangle = \cos(\theta)|0\rangle - \sin(\theta)e^{-i\phi}|2\rangle$ and
	two bright states 
	$|B_1\rangle = \frac{1}{\sqrt{2}}[\sin(\theta)|0\rangle + \cos(\theta)e^{-i\phi}|2\rangle + |1\rangle] $ and
	$|B_2\rangle = \frac{1}{\sqrt{2}}[\sin(\theta)|0\rangle + \cos(\theta)e^{-i\phi}|2\rangle + |1\rangle] $,
	where 
	$\theta = \arctan\frac{\Omega_p(t)}{\Omega_s(t)}$ and
	$\phi$ is the relative phase between the P- and S-pulses. Without loss of generality, we set $\phi = 0$ in this experiment.                                                                                                  
	$\Omega_p(t)$ ($\Omega_s(t)$) is the envelope of the P-pulse (S-pulse) that drives 
	the transition between $|0\rangle$ ($|1\rangle$) state and $|1\rangle$ ($|2\rangle$) state, as shown in the orange zone in Fig. \ref{fig:FIG1}a. 
	
	In our experiment, we use the Gaussian pulses
	\begin{equation}\label{Eq1}
	\begin{aligned}
	&\Omega_p(t) = \Omega_0 e^{-(t-T/2+\delta \tau)^2/\sigma^2}\\
	&\Omega_s(t) = \Omega_0 e^{-(t-T/2-\delta \tau)^2/\sigma^2}
	\end{aligned}
	\end{equation}
	as the P- and S-pulses, respectively, as denoted in Fig. \ref{fig:FIG1}\textbf{b},
	where $\Omega_0$ is the Gaussian pulse amplitude,
	$T$ is the total evolution time,
	$\delta \tau = 	T/11$
	is the separation time between the two pulses, and 
	$2\sigma=T/6$
	is the full-width at half-maximum of the pulse. To accelerate this STIRAP procedure while maintaining high fidelity, we modify the original CD driving approach in two steps. First, we re-construct the CD driving term to avoid two-photon transition. Second, we optimize the driving parameters using the CMA-ES routine. The Hamiltonian of the CD driving term is written as $H_{cd} = -i \Omega_{cd} |0\rangle \langle2|+ h.c.$, where 	$\Omega_{cd}(t)=\frac{\dot{\Omega}_p(t)\Omega_s(t) - \Omega_p(t) \dot{\Omega}_s(t)}{\Omega_p^2(t) + \Omega_s^2(t)}$. This term introduces the two-photon transition between state $|0\rangle$ and $|2\rangle$ in the transmon, which is troublesome for our routine. To simplify this problem, we modify the  additional term in a new frame (see Methods) to cancel the two-photon transition, creating the coherent control pulse transform as below:
	\begin{equation}\label{Eq2}
	\begin{array}{cc}
	\tilde{\Omega}_p(t) = \Omega_p(t) - 2\dot{\zeta}(t) \\
	\tilde{\Omega}_s(t) = \sqrt{\Omega_p(t) + 4\Omega^2_{cd}(t)} \\
	\end{array},
	\end{equation}
	where $\zeta(t) = \arctan{\frac{2\Omega_{cd}(t)}{\Omega_p(t)}}$.

	\begin{figure}
		\begin{minipage}[b]{0.5\textwidth}
			\centering
			\includegraphics[width=8.5cm]{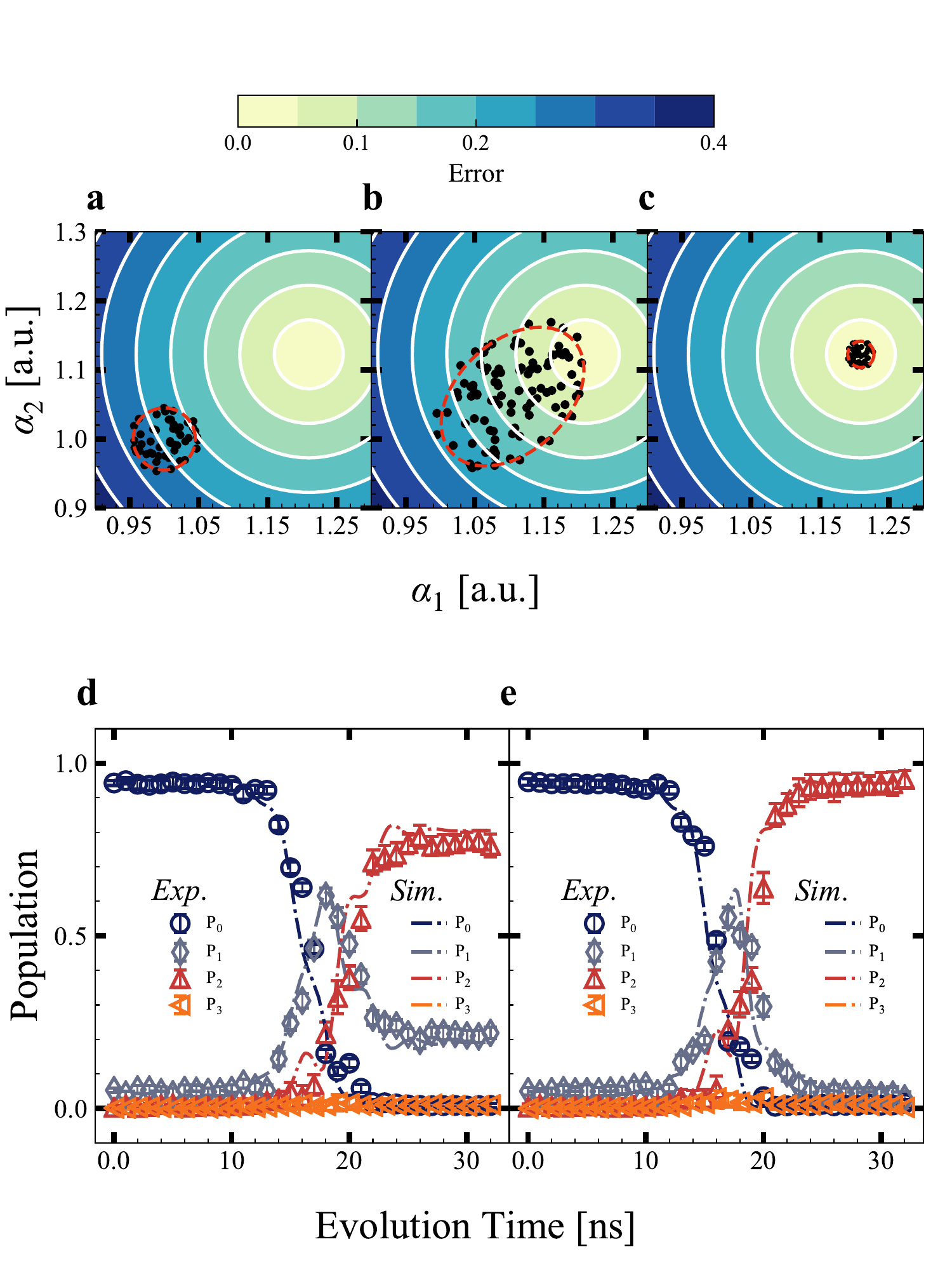}
		\end{minipage}
		\caption{
			\textbf{State transfer control.}
			\textbf{a-c}$\,$The CMA-ES based optimization process in the amplitude subspace of the parameter space.
			\textbf{a}$\,$Initial parameters.  
			\textbf{b}$\,$Intermediate generation parameters.   
			\textbf{c}$\,$Final optimal parameters.
			\textbf{d-e}$\,$State transfer control simulation (line data) and experiment results (point data) 
			with 32 ns manipulation time. 
			\textbf{d} Transfer the initial state (state $|0\rangle$) to the target state (state $|2\rangle$) by STIRSAP.
			The fidelity of transfer control is 
			$0.900 \pm 0.006$ and the fidelity in simulation is $0.919$.
			\textbf{e} Transfer the initial state (state $|0\rangle$) to the target state (state $|2\rangle$) by STIRSAP-Opt.
			The fidelity of transfer control is 
			$0.996 \pm 0.005$, while it is $0.999$ in simulation.
			\label{fig:FIG2}
		}
	\end{figure}
	\begin{figure}
		\begin{minipage}[b]{0.5\textwidth}
			\centering
			\includegraphics[width=8.5cm]{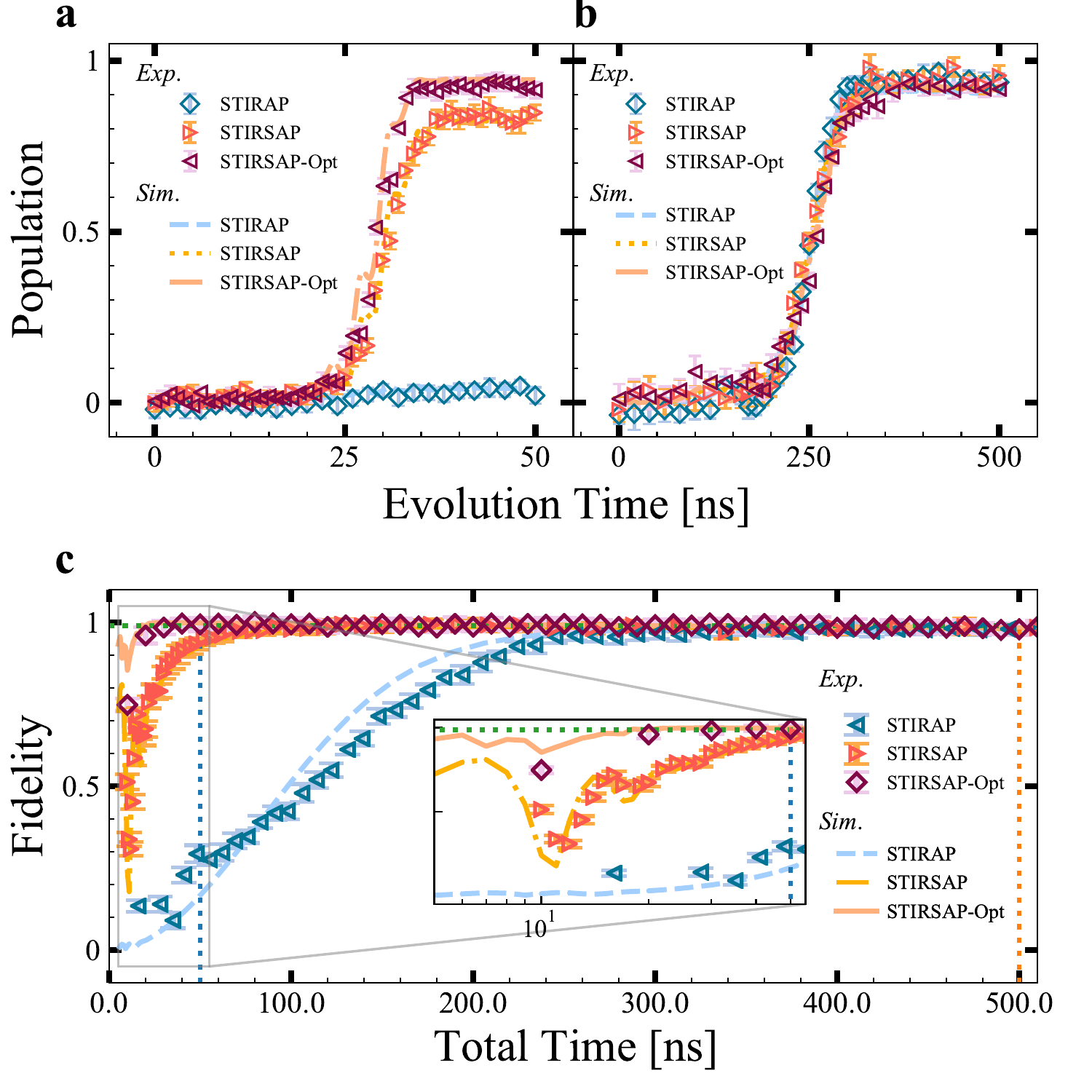}
		\end{minipage}
		\caption{
			\textbf{The fidelity of state transfer control with evolution total time.}
			\textbf{a}$\,$The evolution of the target state ($|2\rangle$ states) population
			is within 50 ns (here, $T = T_0$, $T_0 = 2\pi/\Omega_0$ and $\Omega_0 = 2\pi \times 20$ MHz). 
			The fidelities of transfer control are 
			$0.276 \pm 0.017, 0.935 \pm 0.018, 0.994 \pm 0.021$ 
			for STIRAP, STIRSAP, STIRSAP-Opt, respectively,
			and in simulation, they are $0.179$, $0.932$, $0.998$.
			For $T = 10T_0$ in \textbf{b}, however, it means that the conventional adiabatic approximation condition is satisfied. 
			The fidelities of transfer control are  
			$0.982 \pm 0.013$, $0.984 \pm 0.011$, $0.986 \pm 0.020$, respectively.
			The fidelities in simulation are $0.981$, $0.980$, $0.981$, respectively.
			\textbf{c}$\,$Measured (point data) and simulated (line data) fidelity of transfer control as a function of
			the total time with reference driving amplitude $\Omega_0$ 
			using STIRAP, STIRSAP, and STIRSAP-Opt, respectively.
			The two dotted lines at 50ns and 500 ns mark the moments when the total time 
			is the time $T$ in \textbf{a} and \textbf{b}, respectively.
			The inset image zooms in the part of $T \in [5, 55]$ ns to more intuitively compare the differences between STIRAP, STIRSAP and STIRSAP-Opt.			
			\label{fig:FIG3}
		}
	\end{figure} 
	\begin{figure*}
		\begin{minipage}[b]{1.\textwidth}
			\centering
			\includegraphics[width=18cm]{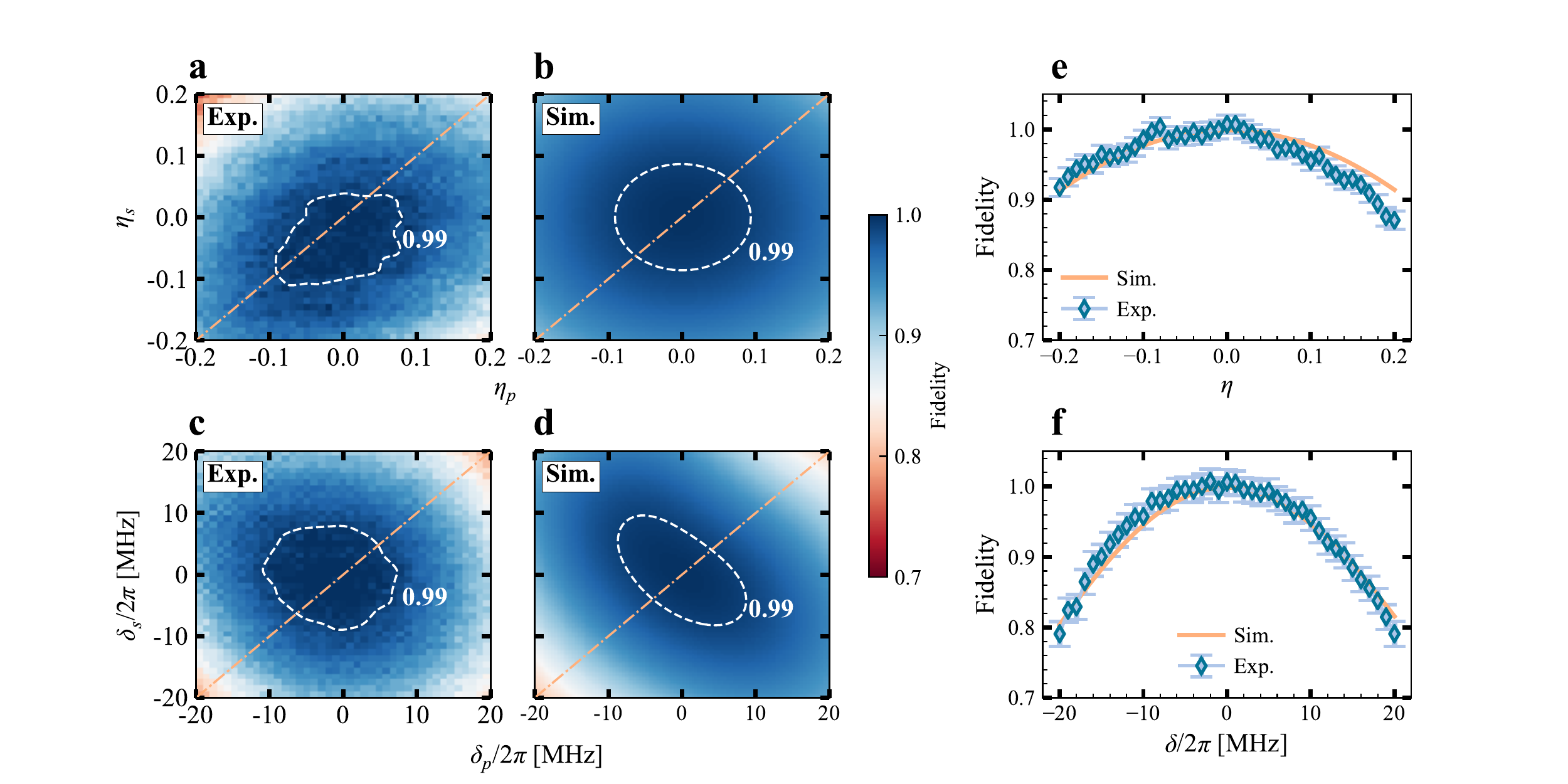}
		\end{minipage}
		\caption{
			\textbf{Robustness.}
			\textbf{a} and \textbf{c} are the robustness of state transfer control against errors 
			in the driving pulse amplitude and detuning for STIRSAP-Opt, respectively,
			and  \textbf{b} and \textbf{d} are respective theoretical simulations.
			\textbf{e} and \textbf{f} are details of the transfer fidelity 
			along the anti-diagonal of \textbf{a} and \textbf{c}, respectively. 
			The points (lines) are experimental (simulation) results.
			\label{fig:FIG4}
		}
	\end{figure*}

	\textit{Optimization of driving parameters.} Now the system Hamiltonian with the STA approach becomes
	\begin{equation}\label{Eq3}
	\begin{aligned}
	H = &\sum_n \omega_n|n \rangle \langle n| \\
	& + \sum_{k=p,s} \sum_{j} [\tilde{\Omega}_k(t) \cos{(\omega_k t + \phi_k)} \sqrt{j}|j-1\rangle \langle j| + h.c.]
	\end{aligned}
	\end{equation}
	where index $j=1, 2, 3$, $n = 0, 1, 2, 3$ (considering the lowest four levels in the transmon) are the labels of energy eigenstates,
	$\phi_k$ is the initial phase of the P-pulse ($k=p$) or the S-pulse ($k=s$), and 
	$\omega_p$ ($\omega_s$) is the P-pulse (S-pulse) frequency. Notice that they can also drive other energy level transitions with certain detuning, which depends on the specific parameters of the transmon. Therefore, the problem we are dealing with is similar to the two-level system population leakage caused by weak anharmonicity, where we have to consider the effect of the spurious coupling terms 
	and the Stark shift of the energy levels in the manipulation. However, the situation here is more complicated due to more energy levels being involved.
	
	Here, we introduce the CMA-ES into STIRSAP  to optimize the driving pulses and achieve 
	high-fidelity state transfer control, which we call STIRSAP-Opt.
	For simplicity, we mainly improve the fidelity by optimizing 
	the amplitudes and detunings of driving pulses. 
	The driving pulses change from Fig. \ref{fig:FIG1}\textbf{c} to Fig. \ref{fig:FIG1}\textbf{d}, which can be written as
	\begin{equation}\label{Eq4}
	\begin{aligned}
	&\Omega_{p}(t) = \alpha_p \tilde{\Omega}_{p}(t), \quad \Delta_{p}(t) = \beta_p\\
	&\Omega_{s}(t) = \alpha_s \tilde{\Omega}_{s}(t), \quad \Delta_{s}(t) = \beta_s
	\end{aligned}
	\end{equation}
	where $\alpha_p$ ($\alpha_s$) and $\beta_p$ ($\beta_s$) represent 
	the amplitude coefficient and detuning of the P-pulse (S-pulse) to be optimized, respectively. 
	We select a section in the parameter space (the amplitudes and the detunings form a four-dimensional parameter space in optimization) to visualize the optimization process. 
	After the initial parameters are set, new candidate solutions (the black solid points) are generated 
	by variations around the initial point, and the dotted line shows the distribution of the dots, 
	as depicted in Fig. \ref{fig:FIG2}\textbf{a}.
	
	In order to optimize parameters using CMA-ES, one can introduce the fidelity of state transfer control\cite{nielsen2002quantum}
	\begin{equation}\label{Eq5}
		\mathcal{F} = \rm{Tr}\sqrt{\sqrt{\rho_{ideal}} \rho_{exp} \sqrt{\rho_{ideal}}}
	\end{equation}
	and have a cost function
	\begin{equation}\label{Eq6}
	\mathcal{C} = 1-\mathcal{F}.
	\end{equation}
	Here, $\rho_{exp}$ is the qudit density matrix at $t = T$, and $\rho_{ideal}$ is the target state we want. 
	In every iteration of the CMA-ES based on \eqref{Eq6}, some dots (individuals) are eliminated, and some dots are selected to 
	become the new candidates (parents) in the next generation.
	Fig. \ref{fig:FIG2}\textbf{b} shows the intermediate generation, and in Fig. \ref{fig:FIG2}\textbf{c}, the optimization converges 
	and gives a global optimal parameter set which will be used in STIRSAP-Opt.
	Given our experimental conditions, to show that STIRSAP-Opt can also achieve high fidelity in a very short period of time, we set $T=32$ ns 
	and Rabi amplitude $\Omega_0 = 30$ MHz to demonstrate our routine.
	According to Eq. \eqref{Eq1}, Eq. \eqref{Eq2}, and Eq. \eqref{Eq4}, we get the driving pulses 
	of three passages, shown in Fig. \textbf{1b-1d}.
	Fig. \ref{fig:FIG2}\textbf{d} and Fig. \ref{fig:FIG2}\textbf{e} show the state transfer process following STIRSAP and STIRSAP-Opt. 
	In these two cases, judging by the fidelity of state transfer and reduction of the population of the intermediate state $|1\rangle$, we find that STIRSAP-Opt has better performance. 
	It is worth specifying that due to thermal excitation, the residue of state $|1\rangle$ exists
	during the entire procedure (More details see Supplementary Note 1). 
	Within the short transfer time (32 ns), the STIRSAP performance is mainly limited by the leakage on a non-computational basis, whose fidelity is $0.900 \pm 0.006$.
	STIRSAP-Opt, however, can speed up the passage with rigorous leakage suppression with a fidelity of $0.996 \pm 0.005$
	(In this article, the error bars indicate the $95\%$ confidence interval).
	Meanwhile, based on Eq. \eqref{Eq3}, we also get 
	the evolving population of states over time and calculate
	the corresponding fidelities are $0.919$ and $0.999$ for STIRSAP and STIRSAP-Opt, respectively,
	and they are in good agreement with the experimental results.
	
	Next, we further study the speed-up of STIRSAP-Opt compared with STIRAP and STIRSAP.
	Without loss of generality, Rabi amplitude $\Omega_0$ is fixed as $20$ MHz.
	We perform STIRAP, STIRSAP, and STIRSAP-Opt with different procedure time $T$, with the respective fidelities shown in Fig. \ref{fig:FIG3}\textbf{c}. 
	The green horizontal dotted line indicates the threshold of fidelity at 0.99. Obviously, the shorter the time, the larger the advantage of our approach.  Here we emphasize two extreme conditions, which are $T=50$ ns and $T=500$ ns (as vertical dotted line denoted in Fig. \ref{fig:FIG3}\textbf{c}). 
	When $T=500$ ns, the performance of all approaches is good, and the fidelities of STIRAP, STIRSAP, and STIRSAP-Opt are  
	$0.982 \pm 0.013$, $0.984 \pm 0.011$, and $0.986 \pm 0.020$,
	while the simulation results are 
	$0.981$, $0.980$, and $0.981$, respectively.
	At $T=50$ ns, the fidelities become 
	$0.276 \pm 0.017$, $0.935 \pm 0.018$, and $0.994 \pm 0.021$,
	and the results in simulation are 
	$0.179$, $0.932$, and $0.998$, respectively,
	proving that our method has significant advantage.
	We also compare the population of $|2\rangle$ in the three passages in Fig. \ref{fig:FIG3}\textbf{a} and Fig. \ref{fig:FIG3}\textbf{b}, and the results agree with the fidelity results.
	Here, we point out that the errors at T = 500 ns mainly come from the influences of decoherence. 
	The performance of STIRSAP-Opt without decoherence sees Supplementary Note 2.
	
	Furthermore, we have experimentally verified that the state transfer control 
	based on STIRSAP-Opt has impressive robustness while maintaining high fidelity.
	As shown in Fig. \ref{fig:FIG4}\textbf{a} and Fig. \ref{fig:FIG4}\textbf{c}, we measured the fidelity 
	by changing the P- and S-pulses amplitude and frequency, 
	which are similar to errors of driving pulses in the experiment. 
	As shown in Fig. \ref{fig:FIG2}\textbf{e}, the length of the driving pulses we use is 32 ns,
	and we define driving amplitude error 
	$\eta_k = 1 - \Omega_k/\Omega_{ref,k}$,
	and driving frequency error 
	$\delta_k = \Delta_k - \Delta_{ref,k}$
	where the index $k = p, s$ means the P- and S-pulses, and 
	$\Omega_{ref,k}$ and $\Delta_{ref,k}$ are optimal amplitude and detuning by STIRSAP-Opt, respectively.
	Meanwhile, we use QuTiP\cite{johansson2012qutip, Johansson_2013} to simulate the process under the experimental conditions. 
	The results are shown in Fig. \ref{fig:FIG4}\textbf{b} and Fig. \ref{fig:FIG4}\textbf{d}.
	In the experiment, the amplitude-frequency response of the experimental circuit 
	is nonlinear under different amplitudes and frequencies.
	For short pulses, the effect of these errors may be especially serious due to leakage caused by the weak anharmonicity, leading to the differences between the experiment and the simulation. 
	In order to highlight the details of the differences, 
	we have selected data along the antidiagonal of experimental results (Fig. \ref{fig:FIG4}\textbf{a} and Fig. \ref{fig:FIG4}\textbf{c})
	and simulation results (Fig. \ref{fig:FIG4}\textbf{b} and Fig. \ref{fig:FIG4}\textbf{d}) to plot Fig. \ref{fig:FIG4}\textbf{e} and Fig. \ref{fig:FIG4}\textbf{f}.
	Here, although the pulses are short (32 ns), our experimental results and simulation results are in good agreement.
	
	~\\ 
	\textbf{Discussion.}
	
	In conclusion, we eliminated the two-photon resonance channel to reduce the complexity of manipulation after 
	introducing CD driving into the STIRAP protocol under resonance conditions in superconducting circuits. 
	Considering the weak anharmonicity in the transmon qudit, we combined the optimization algorithm CMA-ES 
	to achieve fast ($32$ ns) and high-fidelity ($0.996 \pm 0.005$) state transfer.
	Our method can be directly applied to quantum manipulation 
	in quantum communication, especially for systems that are composed of multiple qubits or multi-quantum nodes, which have the loss channel or spontaneous emission of the intermediate state.
	Thus, quantum manipulation with high fidelity and fast speed can be realized 
	with intermediate coherent times and under experimental conditions.
	
	~\\ 
	\textbf{Methods.}
	
	\textbf{Theoretical model}
	Based on the description in the text,
	considering a three-level system and after introducing CD drving, 
	we have the Hamiltonian
	\begin{equation}
	H = \frac{1}{2}
	\left(
	\begin{array}{ccc}
	0				&  \Omega_p(t)  &	\Omega_{cd}(t)	\\
	\Omega_p(t)		&  0  			&	\Omega_s(t)		\\
	\Omega_{cd}(t)	&  \Omega_s(t)  &	0	
	\end{array}
	\right)
	\end{equation}
	where
	\begin{equation}
	\Omega_{cd}(t)=\frac{\dot{\Omega}_p(t)\Omega_s(t) - \Omega_p(t) \dot{\Omega}_s(t)}{
		\Omega_p^2(t) + \Omega_s^2(t)}
	\end{equation}
	is the CD driving term to suppress the non-adiabatic emission in evolution.
	The Hamiltonian itself satisfies the intrinsic SU(2) Lie algebra\cite{PhysRevA.94.063411}.
	Using the Gell-Mann matrices, we have
	\begin{equation}
	H = \frac{1}{2}[\Omega_p(t) \lambda_1 + \Omega_s(t) \lambda_6 - 2\Omega_{cd}(t) \lambda_5],
	\end{equation}
	where
	\begin{equation}
	\lambda_1 = 
	\left(
	\begin{array}{ccc}
	0 & 1 & 0 \\
	1 & 0 & 0 \\
	0 & 0 & 0 \\
	\end{array}
	\right),
	\lambda_5 = 
	\left(
	\begin{array}{ccc}
	0 & 0 & -i \\
	0 & 0 & 0  \\
	i & 0 & 0  \\
	\end{array}
	\right),
	\lambda_6 = 
	\left(
	\begin{array}{ccc}
	0 & 0 & 0 \\
	0 & 0 & 1 \\
	0 & 1 & 0 \\
	\end{array}
	\right).
	\end{equation}
	Therefore, when $U(t) = e^{-i\zeta(t)\lambda_6}$ is introduced for unitary transformation, 
	we can obtain the Hamiltonian 
	\begin{equation}
	H = \frac{1}{2}[\tilde{\Omega}_p(t) \lambda_1 + \tilde{\Omega}_s(t) \lambda_6 - 2\tilde{\Omega}_{cd}(t) \lambda_5]
	\end{equation}
	Suppose $\tilde{\Omega}_{cd}=0$, 
	we get $\zeta(t)$. 
	Under this condition, the initial state and the final state are not directly coupled, 
	and we can get the coherent control pulses based on Eq. \eqref{Eq2}. 
	When the evolution time becomes short, 
	the second-order rotating wave approximation fails in systems with weak anharmonicity, 
	and the Hamiltonian becomes Eq. \eqref{Eq3}.
	We consider four levels and define a unitary matrix 
	$U = \sum_n e^{-i\theta_n}|n\rangle \langle n|$
	(where $\theta_n = \omega_n t$ and $\omega_p$($\omega_s$) is the P-pulse (S-pulse) driving frequency).
	We have
	\begin{equation}
	\begin{aligned}
	H = &\frac{1}{2} [
	\tilde{\Omega}_p(t) 					   +  \frac{1}{\sqrt{2}}\tilde{\Omega}_s(t) e^{-i\delta_p} 		|0\rangle \langle 1| \\
	& + \sqrt{2} \tilde{\Omega}_p(t) e^{i\delta_s} +  \tilde{\Omega}_s(t)] 										|1\rangle \langle 2| \\
	& + \sqrt{3} \tilde{\Omega}_p(t) e^{i\delta_1} +  \sqrt{\frac{3}{2}}\tilde{\Omega}_s(t) e^{i\delta_2}] |2\rangle \langle 3| + h.c.]
	\end{aligned}
	\end{equation}
	We set $\omega_0 = 0$ GHz, and
	$\delta_p = (\omega_1 - \omega_p)t + \phi_p$,
	$\delta_s = (\omega_2 - \omega_1 - \omega_s)t + \phi_s$,
	$\delta_1 = (\omega_1 - \omega_s)t + \phi_s$,
	and $\delta_2 = (\omega_2 - \omega_1 - \omega_p)t + \phi_p$.
	As shown in the formula above and in Fig. \ref{fig:FIG1}\textbf{a}, the fidelity of state transfer is low 
	because there are coupling terms between other unnecessary energy levels.

	~\\ 
	\textbf{Data availability}.
	Data that support the findings of this study are available from the corresponding author upon reasonable request.
	
	~\\ 
	\textbf{Code availability}.
	The codes developed for the simulations of this study are available from the corresponding author upon reasonable request.

	~\\
	\textbf{Acknowledgement.}
	This work was supported by the NKRDP of China (Grant No. 2016YFA0301802), NSFC (Grants No. 11504165, No. 11474152, No. 12074179 and No. 61521001), and the Young Fund of Jiangsu Natural Science Foundation of China (Grant No. BK20180750).
	
	\section{references}

\end{document}